\documentclass[10pt,twocolumn,letterpaper]{article}

\usepackage{cvpr}
\usepackage{times}
\usepackage{epsfig}
\usepackage{graphicx}
\usepackage{amsmath}
\usepackage{amssymb}
\usepackage{multirow}
\newcommand{\ka}[1]{\textcolor{black}{#1}}

\usepackage[pagebackref=true,breaklinks=true,letterpaper=true,colorlinks,bookmarks=false]{}

 \cvprfinalcopy 

\ifcvprfinal\fi
\begin{document}

\title{Natural Disasters Detection in Social Media and Satellite imagery: a survey}

\author{Naina Said\\
DCSE \\ University of Engg and Tech. Peshawar,  Pakistan \\
{\tt\small ahmadka@tcd.ie}
\and
Kashif Ahmad \\
University of Trento, Italy\\
{\tt\small kashif.ahmad@unitn.it}
\and
Michael Riegler\\
Simula Metropolitan Center for Digital Engineering\\ University of Oslo, Norway\\
{\tt\small michael@simula.no}
\and
Konstantin Pogorelov\\
Simula Research Laboratory \\University of Oslo, Norway\\
{\tt\small konstantin@simula.no}
\and
Laiq Hassan\\
DCSE\\ University of Engg and Tech. Peshawar, Pakistan\\
{\tt\small laiqhasan@gmail.com}
\and
Nasir Ahmad\\
DCSE\\ University of Engg and Tech. Peshawar, Pakistan\\
{\tt\small n.ahmad@uetpeshawar.edu.pk}
\and
Nicola Conci\\
University of Trento, Italy\\
{\tt\small nicola.conci@unitn.it}
}

\maketitle

\begin{abstract}
The analysis of natural disaster-related multimedia content got great attention in recent years. Being one of the most important sources of information, social media have been crawled over the years to collect and analyze disaster-related multimedia content. Satellite imagery has also been widely explored for disasters analysis. In this paper, we survey the existing literature on disaster detection and analysis of the retrieved information from social media and satellites. 
Literature on disaster detection and analysis of related multimedia content on the basis of the nature of the content can be categorized into three groups, namely (i) disaster detection in text; (ii) analysis of disaster-related visual content from social media; and (iii) disaster detection in satellite imagery. We extensively review different approaches proposed in these three domains. Furthermore, we also review benchmarking datasets available for the evaluation of disaster detection frameworks. Moreover, we provide a detailed discussion on the insights obtained from the literature review, and identify future trends and challenges, which will provide an important starting point for the researchers in the field.
\end{abstract}

\section{Introduction}
Natural disasters caused by natural processes, such as thunder storms, wildfires, earthquakes and floods, may lead to significant losses in terms of property and human lives. Gathering information about the damage caused by a natural disaster in time is very crucial and may help in mitigating the loss, and faster recovery. 
However, gathering such information is a challenging task due to a number of factors. As an example, situations have been reported where news agencies were unable to provide information about natural disasters in time or at all due to lack of reporters in the area. \cite{stelter2008citizen,yin2012using,ahmad2017jord}. 

Social media emerged as an important source of communication and dissemination of information in emergency situations \cite{ahmad2017jord2,ahmad2018social}. Under such circumstances, inferring disaster events through information available in social media has been an area of interest for the researchers \cite{ahmad2017jord,kaplan2010users,houston2015social}, with particular focus on detection, analysis and summarization of disaster-related multimedia content. Satellite data has also been widely used to analyze the impact of natural disasters on the surface of the earth. In literature, various solutions have been proposed to analyze disaster-related information obtained from social media and satellite in the form of text and images with diverse classification and feature extraction strategies. 

In this paper we provide a detailed analysis of how social media can play a vital role in disaster situations in terms of communication and dissemination of news along with a detailed survey of different approaches proposed for disaster events detection, summarization and filtering of the retrieved information from social media and satellites. We also discuss the current trends, challenges and future directions of research. Moreover, We present a detailed survey of publicly available benchmark datasets in the domain.

The rest of the paper is organized as follows: In Section 2, we discuss the key research challenges in disaster analysis in different domains. Section 3 provides a detailed review of the state-of-the-art approaches for disaster events detection in social media. Section 4 describes literature on disaster detection in satellite imagery. Section 5 reports the current trends in disaster analysis. Section 6 provides the details of the benchmarking datasets available for the evaluation of the proposed solutions in Twitter's text, images from social media and satellite imagery along with databases providing other statistics of natural disasters occurred worldwide. Section 7 draws some concluding remarks and discusses future directions of research on the subject.

\section{Open/Key Research Challenges}

\ka{Although social media and satellite imagery have been proven very efficient in disaster analysis, there are several challenges associated with the use of social media and satellite imagery in general and specifically in disaster analysis. Processing social media content to obtain relevant information, generally, involves challenges of collecting, handling and analyzing a diversified set of information, including textual and visual content, from different social media platforms. In this section, we identify some open research challenges in this domain.}

\ka{The relevance and authenticity of content shared via social media are among the main challenges associated with the disaster analysis in social media. To obtain relevant information, the majority of the works/applications rely on content based analysis of retrieved information as detailed in Section 3. However, the existing literature on disaster analysis lacks in dealing with the authenticity of the content, and additional measures need to be taken to check the authenticity of news and other information shared in social media. 
Moreover, the domain also lacks of public datasets, e.g., for Twitter most of the works \cite{shekhar2015disaster,to2017identifying,klein2013emergency,truong2014identifying} use self-collected datasets. 
Moreover, the datasets are often not large enough in terms of total number of images and types of the natural disaster events they cover. }

\begin{figure}[t!]
	\label{img-nasa}
	\begin{center}
		\includegraphics[width=.98\linewidth]{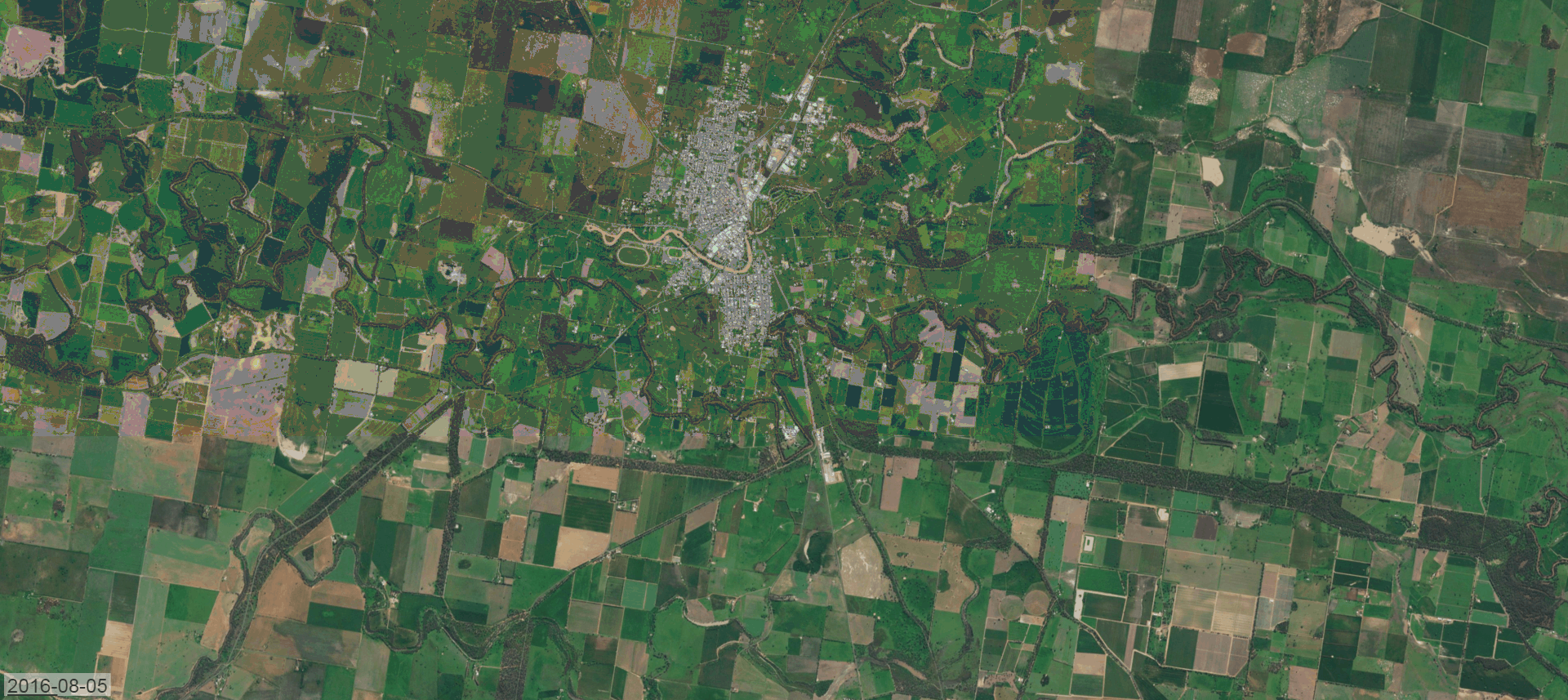}
		\caption[A satellite image of floods in New South Wales, Australia. Based on the image, though the floods can be observed, it is almost impossible to give a clear statement about its impact on environment and society (image from planet)]{A satellite image of floods in New South Wales, Australia. Based on the image, though the floods can be observed, it is almost impossible to give a clear statement about its impact on environment and society (image from plane \url{https://www.planet.com}).
		}
	\end{center}
	\vspace{-0.3cm}
\end{figure}

\ka{Remote-sensed data also come with several challenges. Satellite images have a low temporal frequency, and, more importantly, they only give a bird's-eye view of an event \cite{paul2009new}. For example, as shown in Figure \ref{img-nasa}, floods can be detected in the satellite image of New South Wales, Australia, taken from Planet 4-band satellite\footnote{https://www.planet.com}, but this information does not reveal the impact the flood had on people's life. In such situations, social media information can be used to enrich satellite imagery by providing users with a combined view of satellite imagery and social media information as discussed in Section 5. Moreover, sometimes satellites may not provide a clear view of the ground due to clouds, lighting conditions, vegetation and even processing errors \footnote{\url{https://sites.google.com/site/satelliteimagery/home/satellite-imagery-pros-cons}}. }


\ka{Based on the above discussion, the key research challenges are:}

\begin{itemize}
    \item \ka{The relevance and authenticity of content shared via social media is always a big challenge for applications aiming disaster detection and analysis of disaster-related data available in social media.}
        \item \ka{The domain lacks in large-scale annotated datasets for the training and evaluation of machine learning techniques for disaster analysis in Twitter text and images from social media and satellites.}
            \item \ka{Satellite imagery is not frequently available due to the low temporal frequency of satellite imagery. Moreover, the quality of the satellite imagery may be affected by several factors.}
\end{itemize}

In the next sections we describe how the literature approaches the challenges associated with retrieval and content analysis of information obtained from social media and satellites. We also describe the methods proposed to enrich satellite imagery with social media information. 

\section{Disaster Detection in Social Media}
Disaster-related data captured by users and shared via social media platforms are usually available either in textual or visual form. Based on the type of media, we divide disaster detection in social media into two categories, namely (i) disaster detection in Twitter text and (ii) disaster detection in visual content from social media. 

\subsection{Disaster Detection in Twitter Text}

The literature suggests that Twitter has been widely utilized for communication and dissemination of information in emergency situations in general and particularly during natural disasters \cite{castillo2016big,brouwer2017probabilistic,atefeh2015survey}. For instance, Murthy et al. \cite{murthy2013twitter} provide a detailed analysis of the important role that Twitter has played as a source of communication and information during the floods in Pakistan in 2010. Takahashi et al. \cite{takahashi2015communicating} analyze the use of Twitter before and after the typhoon in Philippines, modeling external factors (time of use and geographic location information) and internal factors (stakeholders and their social media usage). 

In \cite{shekhar2015disaster}, a system has been proposed to utilize users' captured information for the analysis and estimation of the damage, and to analyze the sentiments and level of distress of the users of the system as well as the people affected by the disasters. The proposed system consists of three main modules: data extraction, sorting and analysis. The data extraction/collection and sorting modules are responsible for data collection and pre-processing (cleaning the retrieved data). The third module analyzes the data to extract useful information about the disaster scope, distribution, Geo tagging, occurrence frequency and sentiment rating. In order to extract the location of the disaster from the tweets, the authors use a geo-tagged filter, while K-Nearest Neighbour (K-NN) algorithm has been adopted for the disaster distribution analysis. 
The approach proposed in \cite{to2017identifying} relies on the use of matching keywords and hashtags for the identification of relevant messages from the social media streams. Moreover, the authors also provide a comparative analysis of their proposed matching keywords approach against a learning-based system. In order to evaluate the performance of the two methods, a five step approach has been proposed. These steps are: (i) removal of spam from the data, (ii) mapping of the data to affected and unaffected regions, (iii) filtering of irrelevant tweets, (vi) sentiment analysis and finally (v) visualization of the data. Three types of natural disasters (floods, earthquakes and wildfires) are considered for the comparative study. Based on the experimental results, the authors conclude that the learning-based technique proved to collect a higher number of relevant tweets as compared to the matching-based classification. 
In \cite{klein2013emergency}, Twitter text streams are filtered based on the seriousness of the content analyzed through Natural Language Processing (NLP) techniques. The proposed system extracts facts using NLP and uses an event identification scheme for grouping posts. The basic insight behind the proposed system is to group the posts providing useful information about the underlying disaster events. To this aim, the authors rely on a state-of-the-art library, namely Named Entity Recognizer (NER) \cite{stanford2010stanford}, to extract emergency facts, followed by some pre-processing steps to clean the data by removing stop words. The matchmaking approach is then used for assigning the posts to specific emergency clusters. 

Truong et al. \cite{truong2014identifying} rely on a Bayesian approach for the identification and classification of disaster-related tweets to differentiate hurricanes and sand storms tweets from the conversational ones. 
The proposed system uses an effective set of features, which are feed into a Bayes classifier. 
The designed feature set consists of only nine features unlike the bag-of-words (BoW) model \cite{zhang2010understanding}, which has more than 3,000 features. 
The system with reduced features set is more suitable for the hand-held devices with less computational power. 
Cresci et al. \cite{cresci2015crisis} address the limitations of the existing conventional approaches used for event detection in social media, and propose a crisis mapping system targeting the problem from two different perspectives. For the detection/extraction of disasters tweets, the proposed system relies on Support Vector Machines (SVMs) trained on a wide set of linguistic features including raw and lexical textual features, morpho-syntactic features, syntactic features, lexical expansion features and sentiment analysis features \cite{cresci2015linguistically}. To determine the location and origin of the messages, a novel geo-parsing technique has been proposed. Jing et al. \cite{jing2016flood} introduce the concept of ontology for disaster detection in Twitter streams. The concept enables the linkage of text and image analysis at an hierarchical level. The ontology, which refers to defining a set of primitives like classes or attributes and the relationship among the members of the classes, has been a useful concept and extensively used in other various recognition tasks including object recognition \cite{arvor2013advances}, character recognition \cite{eutamene2011new} and emotion recognition \cite{zhang2013ontology}. The ontology considered in the study illustrates that a flood-related image may contain both water and people. As a first step, the proposed system analyzes Twitter's text. To this aim, a flood events-related corpus is obtained from Twitter. As a result of this step, information like event location, time, description and URLs of the images are obtained. The URLs help in the retrieval of relevant images, which are then used for training the recognition system according to the BOW model. 

There are also some works that, in contrast to conventional approaches attempting to classify messages related to specific disasters, deal with the identification of social media messages on a wide range of possible emergency situations due to both technological and natural disasters. 
For instance, in \cite{pekar2016detecting}, an ensemble framework has been proposed to deal with  heterogeneity issues associated with retrieving information from a wide range of disaster events. The basic motivation for the framework comes from the analysis that messages related to different emergencies tend to have different feature distribution, and, thus a single classifier trained for specific type of disaster is most likely to fail in case of a different event. The system relies on a combination of multiple classifiers each one aiming to a specific type of emergency situation along with a semi supervised generic classifier to detect unknown/new events. The concept of the generic classification, also known as co-training \cite{blum1998combining}, has also been used for domain adaptation in other applications \cite{steedman2003bootstrapping}. It makes use of a large collection of unlabelled messages to acquire additional training data. The unlabeled data is collected by crawling Twitter API using twenty four different types of mass emergencies. In addition, several techniques, such as text normalization and stop word removal, have been used in the pre-processing step for cleaning the collected tweets.  For the classification purposes, maximum entropy and linear SVMs are used based on the proven better performance in \cite{pekar2016selecting}. The authors in \cite{parilla2014automatic} propose an intelligent system to automatically detect tweets related to Habagat flooding incident, which happened in Manila in 2012. SVMs and Naive Bayes classifiers are then trained on textual features extracted through BoW model. 

With the increasing mass production and dissemination of data over social media, the reliability of the information being shared is often under criticism. 
To this aim, in \cite{feng2018extraction}, deep learning methods are applied on user generated text and images shared through Twitter to retrieve high quality eyewitnesses of rainfall and flooding events. The proposed system detects flood-related tweets through spatio-temporal clustering \cite{kisilevich2009spatio}. Earle et al. \cite{earle2012twitter} present an earthquake detection system solely relying on Twitter's text streams. The earthquake events are detected using short term average, long term average algorithm (STA/LTA) algorithm, commonly used in seismology. As a first step, a tweet-frequency time series also called tweetgram is generated on the basis of the collected tweets binned into five-second windows and later normalized to tweets-per-minute. Next, a characteristic function responsible for the event decision is extracted from the tweetgram. The proposed system is evaluated on a collection of tweets containing words like "earthquake", "gempa", "temblor", "terremoto" and "sismo".
Crooks et al. \cite{crooks2013earthquake} analyze the temporal and spatial characteristics of Twitter feed to detect an earthquake event in Twitter streams. Ashktorab et al. \cite{ashktorab2014tweedr} propose Tweedr, a text mining tool for the extraction of useful information from tweets during natural disasters. Tweedr consists of three main components, namely classification, clustering and extraction. The goal of the classification part is to identify the tweets reporting damage or casualties. Clustering helps to merge tweets related to similar events. During the extraction phase, tokens and phrases reporting useful information about different classes, such as infrastructure damage, damage types and casualties are extracted. In total, twelve crisis events occurred in North America since 2006 are considered for the evaluation of the proposed system. For classification purposes, the study evaluates several classification algorithms, and finds Logistic Regression to be the most reliable on several evaluation metrics. For extraction, Conditional Random Fields (CRF) with several different types of features have been proposed. 
As far as the clustering component of the proposed system is concerned, bloom filters \cite{gupta2011enhanced} and SimHash \cite{breitinger2013frash} algorithms are used to form clusters of the events. In \cite{middleton2014real}, Middleton et al. propose a real-time crisis mapping system, able to geo-parse tweets about a specific area of interest in real-time by using geo-spatial data obtained in an offline phase. 
In the offline phase, OpenStreetMaps\footnote{\url{https://www.openstreetmap.org/\#map=7/53.465/-8.240}} 
and GooglePlaces API\footnote{\url{https://cloud.google.com/maps-platform/places/}} are used to obtain street-level and building information, which are then stored in a database for later use. 
In the real time phase, a crawler is used to collect and save crisis-related tweets using specific keywords. 

The selection of informative features also plays an important role in classification of text streams. In case of natural disaster detection in Twitter text streams, Pekar et al. \cite{pekar2016selecting} provide a detailed analysis and comparison of different types of textual features. In total, five different types of features, namely lexical, grammatical, semantic, stylistic and twitter metadata, are evaluated. The evaluation dataset consists of 24,589 tweets out of which 2,193 are labeled to be originated from eye witnesses. The classification goal of the study is to identify whether a given tweet was an eye witness report or not. 
The impact of data heterogeneity is studied by defining two different scenarios for the experiments. In the first scenario, the entire dataset is split randomly into training and test data in proportion of 1 to 9 to ensure that the presence of the same crisis in both test and training sets. This also ensures the feature distribution in both test and training is similar. The second scenario reflects a more realistic case where it is ensured that the tweets related to a crisis instances present in the training data are not present in the test set. The performance is then measured using five classifiers including a SVM, Logistic Regression (LR), Random Forest (RF), Naive Bayes and K-NN. 

Recently, Imran et al. \cite{imran2014aidr} proposed a platform, called AIRD, for automatic detection and classification of disaster-related messages shared by users during crisis events. The system jointly utilizes human intelligence and machine learning for the analysis of a large-scale data at high speed. The system is able to continuously retrieve disaster-related information, and user-defined classification categories are defined simultaneously via crowd-sourcing. The system has been successfully tested for an earthquake event in Pakistan in 2013. The same authors present \cite{vieweg2014integrating} a top-down approach using social media information for rapid assessment of sudden onset disasters. The basic idea behind the system is to help humanitarian agencies in their aid activities. Firstly, the needs of such organizations are investigated and solutions are then proposed accordingly. 

In Table \ref{twitter_summary}, we summarize the characteristics of some of the methods presented in this section. 
\begin{table*}[]
\centering
\caption{Summary of some relevant works in disaster detection in Twitter text in terms of event types, modality (Single, multi-modal) of information, datasets used for the evaluations, and a brief description of the method.}
\label{twitter_summary}
\scalebox{0.69}{
\begin{tabular}{|p{.5cm}|p{2.5cm}|p{1.3cm}|p{1.6cm}|p{5cm}|}
\hline
\multicolumn{1}{|c|}{Refs.} & \multicolumn{1}{c|}{Events} & \multicolumn{1}{c|}{Modality} & \multicolumn{1}{c|}{Dataset}  & \multicolumn{1}{c|}{Method} \\ \hline

\cite{shekhar2015disaster}& Earthquake, droughts, floods and forest fires  & S& Self-collected & The system checks the data for disaster distribution, Geo tagging and occurrence frequency. Location of the disaster is determined using Geo filter tag whereas distribution analysis is done using KNN. \\ \hline

\cite{to2017identifying} & Earthquake, flood and wildfire & S & Self-collected & A five step framework has been proposed to compare the performance of matching based and learning based approach for disaster detection. The steps include spam removal, data mapping to affected and unaffected regions, filtering of tweets, sentiment analysis and visualization. \\ \hline

\cite{klein2013emergency} & Natural (flood, hurricane, fires ,earthquake) and human disasters (epidemics) & S & Self-collected & Makes use of a state-of-the-art library namely Named Entity Recognizer (NER) \cite{stanford2010stanford}, embedded with several NLP techniques, to extract emergency facts. The matchmaking approach is then used for assigning the posts to specific emergency clusters. \\ \hline

\cite{truong2014identifying} & Hurricane & S & Self-collected &Relies on a Bayesian approach for the identification and classification of disaster related tweets. In contrast to \cite{zhang2010understanding}, makes use of an effective reduced set of features consisting of only nine features. The system with reduced features set has been proved suitable for the hand held devices with less power capacity. However, a significant improvement in the performance has been reported by jointly using the newly extracted features with bag of words. \\ \hline

\cite{cresci2015crisis} & Infrastructure or community damages & S & Publicly available Dataset \cite{cresci2015swdm} & Proposes a crisis mapping system relying on SVMs trained on a wide set of linguistic features including raw and lexical text features, morpho-syntactic features, syntactic features, lexical expansion features and sentiment analysis features \cite{cresci2015linguistically}.  \\ \hline

\cite{ashktorab2014tweedr} & 12 Natural disasters events & S &Self-collected & Proposes a text mining tool namely Tweedr, mainly consists of classification, clustering and extraction. The classification component is composed of a combination of several techniques and aims to identify relevant posts. The clustering and extraction components aim to gather tweets on same events and highlight the sentences of the text providing demage information, respectively. \\ \hline

\cite{pekar2016selecting} & 26 mass emergency events & S & CrisisLexT26 \cite{olteanu2015expect} & Proposes and ensemble based framework where several classifiers including SVM, LR, RF, Naive Bayes and KNN, each trained for a specific type of event are combined. Moreover, the performances of different textual features are evaluated for a better representation of twitter posts. In addition, several techniques, such as text normalization and stop word removal, have been used in the pre-processing step for cleaning the collected tweets. \\ \hline

\cite{earle2012twitter} & Earthquakes  & S & self-collected & Relies on well-known concept of seismology, namely short term average, long term average algorithm (STA/LTA) algorithm, for the detection of earthquake events in Twitter's text streams. The system is tested on a collection of tweets  containing specific keywords, such "earthquake", "gempa", "temblor", "terremoto" and "sismo". \\ \hline

\end{tabular}} 
\end{table*}

\subsection{Disaster Detection in images from Social Media}
Similar to Twitter, images shared through social media have also been widely utilized for disaster analysis \cite{ahmad2017jord,ahmad2018social,lagerstrom2016image}. In this regard, the additional information, such as users' tags, geo-location and temporal information, available in the form of meta-data have been proved very effective, both individually and in combination with visual features. Yang et al. \cite{yang2011hierarchical} rely on users' tags associated with Flickr images in combination with visual features. For the representation of visual content, two different types of features, namely HSV low-level color features \cite{sural2002segmentation} and mid-level object features extracted through simultaneous partition and class parameter estimation (SPCPE) \cite{chen1999indexing}, are used. On the other side, textual features are extracted through word-frequency \cite{salton1988term}. In \cite{bischke2017detection}, existing CNN models pre-trained on ImageNet \cite{deng2009imagenet} are used as feature descriptor for the representation of visual information along with the additional information available in the form of meta-data including users' tags, owners and users' information, capturing date and time, as well as geo-location information. Both types of information are evaluated individually and in combination, on floods-related images from social media. For the joint use of the textual and visual information, feature vectors are concatenated to form a single feature vector. 

More recently, Alam et al. \cite{alam2018processing} proposed an image analysis framework by combining human experts and machine learning algorithms for the analysis of social media content during emergency situations. The framework is intended to perform two different tasks including (i) collection and filtering of social media content, and (ii) extraction of actionable information and classification of the disaster-related images. For the collection of images, the authors rely on the publicly available Artificial intelligence for Disaster Response (AIDR) system \cite{imran2014aidr}. Human annotators are then used to annotate the collected images. For the classification purposes, VGGNet-16 \cite{simonyan2014very} is fine-tuned on the collected images. In another work from the same authors\cite{nguyen2017damage}, a detailed analysis of disaster-related images from social media has been provided to determine the level of damage due a disaster. To this aim, a number of existing techniques, such as fine-tuning of existing CNN models and BoW model \cite{zhang2010understanding}, have been evaluated and compared. In \cite{ahmad2018comparative}, authors provide a detailed comparison of different feature extraction and classification techniques on two different datasets with a large number of images from different types of natural disaster events. For feature extraction, two different families of algorithms, namely global and deep features, have been employed. For global features, seven different feature descriptors, namely Colour and Edge Directivity Descriptor (CEDD) \cite{chatzichristofis2008cedd}, Joint Composite Descriptor (JCD) \cite{manjunath2001color}, AutoColor-Correlogram \cite{huang1997image}, Color-Layout (CL) \cite{kasutani2001mpeg}, Edge-Histogram (EH) \cite{won2002efficient}, Pyramid of Histograms of Orientation Gradients (PHOG) \cite{datta2009berkeley}, and Tamura \cite{howarth2004evaluation}, are used from Lire library \cite{lux2016lire}. On the other hand, deep features are extracted through eight different CNN models pre-trained on ImageNet and Places \cite{zhou2014learning} datasets. For classification purposes, the performance of six different classification algorithms, namely SVMs, K-NN, Naive Bayes, Decision Trees, RF and Bayes Network, are evaluated on the complete set of features. Then, a late fusion method is used to combine the classification scores obtained through different combinations of feature extraction and classification algorithms. 

Disaster detection and classification in images from social media has also been part of benchmark challenges. In the next sub-section, we provide a detailed description of the methods proposed for the challenge.

\subsubsection{\ka{Benchmark competitions on disaster analysis in social media}}

\ka{Disaster detection in users' captured images from social media has been introduced as a sub-task in benchmarking initiatives of MediaEval-2017\footnote{http://www.multimediaeval.org/mediaeval2017/} and MediaEval-2018\footnote{http://www.multimediaeval.org/mediaeval2018/}. The initiatives mainly focus on flood events with slight modification in the task. }

In MediaEval-2017 \cite{bischke2017multimedia}, participants were asked to propose solutions for the retrieval of flood-related images from social media. The participants were provided with meta-data along with visual content in the form of images. In order to evaluate the significance of both meta-data and visual information, three mandatory experiments relying on (i) visual features only, (ii) meta-data only and (iii) a combined use of the both types of information were included in the challenge. The meta-data included users' tags along with owner, upload, geo-location and temporal information. The majority of the proposed solutions rely on deep architectures for feature extraction from the images. The framework proposed in \cite{ahmad2017convolutional} relies on existing CNN models, pre-trained on both ImageNet and Places dataset, to extract object and scene-level features from images, respectively. Individual SVMs are then trained on the features extracted with both models, whose classification scores are then combined in a late fusion method. A RF classifier is trained on meta-data. Finally, the classification scores of both classifiers are combined in a late fusion method for the final classification score. A similar approach is adopted in \cite{ahmad2017cnn}. Here, the classification scores, obtained by the classifiers trained on visual features extracted through seven different deep models, are combined with the output of a RF classifier trained on meta-data following three different late fusion techniques: Induced Ordered Weighted Averaging Operators (IOWA) \cite{yager1999induced}, Particle Swarm Optimization (PSO) \cite{shi1999empirical} and average weighting method. 
%
Tkachenko et al. \cite{tkachenko2017wisc} propose a more sophisticated solution for meta-data, which relies on word embedding trained on the entire YFCC100m dataset \cite{thomee2016yfcc100m} along with a machine translation technique for the translation of users' tags into English. For visual features the authors rely on handcrafted visual features, namely CEDD, CL and Gabor \cite{bai2009novel}. For the classification purposes, several algorithms including Logistic Regression (LR) classifier \cite{harrell2001ordinal}, RF, Multi-nominal Naive Bayes and Multi-layer Perceptron \cite{mitra1995fuzzy} are employed. 

The handcrafted visual features provided by the task organizers are also used by Hanif et al. \cite{hanif2017flood}, where visual features extracted through several algorithms including Auto-color correlation, CEDD, EH, Tamura and Gabor are integrated via an early fusion technique, followed by Kernel Discriminant Analysis \cite{hand1982kernel} dimensionality reduction. The same technique is used for the joint use of meta-data and visual features. For the representation of meta-data, Term Frequency Inverse Document Frequency (TF-IDF) \cite{salton1988term} is calculated. For classification purposes, the authors investigated various machine learning techniques, such as RF and SVMs. No significant difference has been observed by combining visual features and meta-data on the test set. However, the authors claim for better performance with the experiments performed using Spectral Regression in combination with Kernel Discriminant Analysis (SRKDA) \cite{cai2007efficient} based fusion of visual and meta-data on the development set. Minh-Son et al. \cite{dao2017domain} propose a domain-specific late fusion method to jointly utilize several types of visual features along with meta-data for the task. In detail, several techniques, such as late fusion, tuning, ensemble learning, object detection and temporal spatial-based event confirmation, are integrated based on the domain specific criteria. For image-based retrieval of flooded images, the study formalizes the task as a problem of ensemble learning and tuning where the visual features are used with supervised learners. 

\ka{In MediaEval-2018 \cite{bischke2018multimediasatellite}, the task focuses on the problem of road passability, and participants were asked to first (i) identify images providing an evidence for passability, and then (ii) differentiate between passable and non-passable roads, if an evidence for passability has been predicted in the first step. The majority of the methods proposed for the task rely on deep architectures with particular focus on existing deep models pre-trained on ImageNet and Places datasets. For instance, Said et al. \cite{nain2018multimediasatellite} propose three different fusion techniques, including an early, late and a hybrid fusion. Moreover, four different deep models (two pre-trained on ImageNet and two on Places dataset) are used as feature descriptors. 
Fen et al. \cite{Feng2018multimediasatellite} rely on multiple deep models. However, all the models are pre-trained on Imagenet only. The additional information including users' tags, textual description of the images, geo-location, temporal and user's information, are used to complement visual features. Textual features are extracted through \textit{fasttext} \cite{bojanowski2017enriching}. }

\ka{Lura et al. \cite{lura2018multimediasatellite} also rely on the joint use of multiple deep models in a framework with double-ended classifier and a compact loss function. The framework aims to solve both tasks in a sequential fashion, where images are first analyzed for evidence of passability and then classified into passable and non-passable classes. The method also relies on data augmentation techniques to increase the training samples. For the textual features, they rely on word embedding initialized with Glove \cite{pennington2014glove}. Moreover, the number of re-tweets and the number of times a tweet has been favoured are also used. In \cite{Armin2018multimediasatellite} ResNet-50 is fine-tuned with two different strategies using different number of epochs. In the first experiment, only the top layer is fine-tuned, while in the second experiment the whole network is fine-tuned on the new set of images. For textual features, tweets are initially translated into English through the Google Translator API; next, BoW descriptors are formed from the text. SVMs are then trained on the extracted features. TF-IDF representation is also employed for the textual features. In \cite{Anastasia2018multimediasatellite}, a framework with two deep architectures (VGGNet) is proposed: initially an image is checked for evidence of passability, then the images providing an evidence of passability are fed into the second model, which classifies them into passable and non-passable classes. Textual features are represented through the \textit{word2vec} method \cite{mikolov2013distributed}. Hanif et al. \cite{Hanif2018multimediasatellite} rely on several features, including CEDD, CL, Fuzzy Color and Texture Histogram (FCTH) \cite{chatzichristofis2008fcth}, EH, JCD and Scalable Color (SC) \cite{albuz2001scalable}, in an ensemble framework. A CNN-based local feature descriptor is also used for the extraction of local features. For textual features, TF-IDF is computed. The textual and visual features are evaluated individually as well as jointly combined in the ensemble framework used for the fusion of visual features.}

Table \ref{single_images_summary} summarizes the main features of the approaches proposed for disaster detection in single images in terms of event types they target, the datasets used for the evaluation, and a brief description of the method.

\begin{table*}[]
\centering
\caption{Summary of some relevant works in disaster detection in single images in terms of event types, modality (single, multi-modal) of the information and the dataset used for the evaluations, and a brief description of the method.}
\label{single_images_summary}
\scalebox{0.69}{
\begin{tabular}{|p{.5cm}|p{2.0cm}|p{.5cm}|p{2cm}|p{6cm}|}
\hline
\multicolumn{1}{|c|}{Refs.} & \multicolumn{1}{c|}{Events} & \multicolumn{1}{c|}{Mod.} & \multicolumn{1}{c|}{Dataset}  & \multicolumn{1}{c|}{Method} \\ \hline

\cite{ahmad2017cnn} & Floods & M & DIRSM \cite{bischke2017detection} & Relies on seven different CNN models pre-trained on ImageNet \cite{deng2009imagenet} and Places datasets \cite{zhou2014learning} for feature extraction. Individual SVMs classifiers are then trained on the features extracted through each models. Also uses three different late fusion methods to combine their classification scores \\ \hline

\cite{ahmad2018comparative} & 8 different types of natural disasters& S &self-collected and DIRSM & Provides a comparative analysis of global and deep features for the analysis of user captured natural disaster related images. For global and deep features 8 and 9 different algorithms are used. Moreover, a comparative analysis of 6 different classification techniques is provided. In addition, a late fusion method is used to combine the classification scores. \\ \hline

\cite{ahmad2017jord2}& 8 different types of natural disasters& M &Self-collected & provides a complete system with query generation and refinement, crawling and filtering the collected data from different platforms of social media. For the content analysis, it relies on deep features with SVM classifier. \\ \hline

\cite{bischke2017detection} & Floods& M & DIRSM  & Uses two different architectures namely DeepSentiBank \cite{chen2014deepsentibank} and X-ResNet \cite{jou2016deep} for feature extraction, which are then concatenated to form a single feature vectors. SVMs are then trained on the for classification purposes. Moreover, textual and visual features are also evaluated individually and in combination using an early fusion method.\\ \hline

\cite{ahmad2017convolutional} & Floods& M & DIRSM & Deep features are extracted through multiple CNN models which are trained on ImageNet and Places datasets. The scores obtained through SVM classifier trained on the features extracted from the individual models are combined. \\ \hline

\cite{tkachenko2017wisc} & Floods& M  & DIRSM & Visual features combination including CEDD, CL and Gabor are used with Logistic Regression Classifier. The visual and metdata features are also combined to improve the accuracy. \\ \hline
\cite{zhao2017retrieving} & Natural disasters events& M & DIRSM & Proposes preprocessing operations like image cropping and test-set pre-filtering. The classification is done using SVM. \\ \hline 

\cite{hanif2017flood} & Floods& M & DIRSM & Relies on Kernel Discriminant Analysis using Spectral Regression. Visual and metadata features are combined to test the performance. \\ \hline 
\cite{dao2017domain} & Floods& M & DIRSM & Relies on a domain specific and late fusion based system. The image retrieval task is formalized as a problem of ensemble learning and tuning. For metadata based retrieval, a Feed Forward Neural Network is used. \\ \hline 

\cite{avgerinakis2017visual} &Floods&  M  & DIRSM & propose Deep Convoutional Neural Networks,DBpedia spotlight and combMax for the retrieval of disaster related social media images. DCNN is adopted by training GoogleNet \cite{szegedy2015going} on 5,055 ImageNet concepts.  \\ \hline

\cite{Hanif2018multimediasatellite} & Floods& M & FCSM \cite{bischke2018multimediasatellite} & relies on several handcrafted visual features as well as CNN-based local features extracted through DELEF \cite{noh2017largescale}, which are combined in an ensemble framework. For textual features relies on TF-IDF. Also reports results of fusion of textual and visual features. \\ \hline

\cite{nain2018multimediasatellite} & Floods& S & FCSM  & Makes use of deep features extracted through several deep models including AlexNet, VGGNet and ResNet, pre-trained on both ImageNet and Places datasets, which are fused using an early, late and hybrid fusion. In the hybrid fusion, the results of both early and late fusion are combined in an additional late fusion achieving the best results among the methods.\\ \hline



\end{tabular}} 
\end{table*}

\section{Disaster Detection in Satellite Imagery}
Over the last decades, satellite imagery has been widely used in a diversified set of applications ranging from agriculture \cite{frolking2002combining,rhee2010monitoring} to forestry \cite{dalponte2008fusion}, target detection \cite{li2011saliency} and regional planning to warfare \cite{campbell2011introduction}. Satellite imagery has also been widely utilized to monitor natural disasters and other adverse events, to analyze their impact on the environment. According to the authors in \cite{joyce2009review}, satellite imagery can be used in all four phases, namely reduction, readiness, response and recovery, of the standard disaster management process. The literature shows that the disaster management systems relying on satellite data mostly focus on the acquisition and pre-processing, change detection and prediction of natural hazards from satellite imagery \cite{gillespie2007assessment}. For instance, Joyce et al. \cite{joyce2009review}, provide a detailed survey of different data types, data acquisition and processing techniques used to map and monitor different types of natural disasters, such as floods and earthquakes, in remote sensed data. Voigt et al. \cite{voigt2007satellite} focus on the services provided by a dedicated disaster management system, the Center for Satellite Crisis Information (ZKI), promoted by the German Aerospace Center (DLR). 
Kerle et al. \cite{kerle2002satellite} investigated the importance and role of satellite imagery in Lahar (mud flow) disaster management. 
Eguchi et al. \cite{eguchi2003resilient} analyze damage to buildings in urban areas due to earthquakes through comparative analysis of satellite imagery of the affected ares captured before and after an earthquake. Jaiswal et al. \cite{jaiswal2002forest} use satellite imagery for the identification and mapping of the fire risk zones, and recording the frequency, at which the zones are affected by the fire.  A similar study is carried out in \cite{youssef2011flash}, where satellite imagery is used for the estimation of the flash flood risk levels of sub-watersheds within the Wadi Feiran basin. In detail, firstly a set of parameters are used to capture the characteristics of the drainage system for flood risks, followed by understanding/highlighting the active zones through a comparison of the effectiveness of the sub-basins. 

Amit et al.~\cite{amit2016analysis} propose a CNN-based framework for the detection of natural disasters in satellite imagery. The proposed CNN model consists of three convolutional and max-pooling layers followed by two fully connected layers. For the evaluation purposes, a dataset is collected covering a sufficient number of satellite image patches from two different types of natural disasters, namely landslides and floods. In~\cite{kamilaris2018disaster}, an existing deep model \cite{simonyan2014very}, pre-trained on ImageNet \cite{deng2009imagenet}, is fine-tuned on aerial photos captured through unmanned aerial vehicles (UAV) during or after different types of natural disasters, namely floods, fires and building collapsed. Another work aiming damage assessment of natural disasters in images taken through UAV has been proposed by Nazr et al. \cite{attari2016nazr}. The adopted network is composed of two components. The first one aims to object localization and the other relies on the FV-CNN \cite{cimpoi2015deep} for the differentiation of different damage-levels. Liu et al.~\cite{liu2016geological} also rely on deep models along with wavelet transformation for the automatic detection of disaster affected areas in satellite imagery. 
Bischke et al. \cite{bischke2016contextual} rely on deep features for the representation of images from 3 different types of natural disasters, namely floods, snow-storm and wildfires. For the extraction  of the deep features, AlexNet \cite{krizhevsky2012imagenet}, GoogleNet \cite{szegedy2015going}, VGGNet \cite{simonyan2014very} and ResNet \cite{he2016deep}, pre-trained on ImageNet are used. SVMs are then trained for the classification of disaster-related images. 

\subsubsection{\ka{Benchmark competitions on Flood detection in Satellite Imagery}}

\ka{Similar to the previous case, flood detection in satellite imagery has also been part of the MediaEval benchmark competitions. In this section, we provide a detailed survey of the approaches proposed for both benchmark challenges.}

In MediaEval 2017 \cite{bischke2017detection}, the task aimed to develop systems that can automatically identify and differentiate flooded and non-flooded regions in satellite imagery. The satellite imagery for the task has been taken from Planet's 4-band satellites \cite{team2016planet}. The imagery contains four channels with RGB, and Near Infrared band information (IR). The majority of the approaches proposed in the response of the challenge rely on deep architectures using all the four channels. Benjamins's et al. \cite{bischke2017detection} treated the task as an image segmentation problem by proposing a deep model \cite{simonyan2014very} composed of a pre-processing phase prior to training of the model using three different strategies. In the pre-processing step all the four components are normalized. For the training stage the authors adopt a modified VGGNet \cite{simonyan2014very} model.
In another work \cite{nogueiradata}, the authors propose four different versions of a network with different number of dilated convolutional layers. All the models are trained with overlapping patches each of size $25\times25$. The same strategy is used in the prediction phase, where the final result is based on the average probabilities of all patches. 

Another interesting solution \cite{ahmad2017cnn} proposed for flood detection in satellite imagery relies on a Generative Adversarial Networks (GANs) \cite{goodfellow2014generative} framework. The approach treats the task as a generative problem, where a GAN originally proposed for the retinal vessel segmentation, namely V-GAN \cite{son2017retinal}, is adopted with slight modifications. In order to generate binary segmentation masks of the flood related images, the top layer is equipped with a threshold mechanism. As an extension of the work, the same authors performed a number of experiments with diverse strategies \cite{ahmad2018social}. The network is modified to support both three-channel (i.e., RGB only) and four-channel RGB+IR geo-image-compatible input images. In the case of four-channel input, both the RGB and IR components are normalized independently. 
According to the authors and the conducted experiments, the IR component may also lead to false positives. 
In another work \cite{fu2017bmc}, deep features are extracted through an existing CNN model, ResNet \cite{he2016deep}. Subsequently a RF classifier is used for the classification of the satellite image patches into flooded and non-flooded regions. 
Tkachenk et al. \cite{tkachenko2017wisc} proposed a solution based on the selection of spectral images with three indices, namely LWI (Land Water Index), NDVI (Normalised Difference Vegetation
Index) and NDWI (Normalised Difference Water Index). For the classification purposes both supervised classification and unsupervised clustering techniques are used. Moreover, K-Means clustering is used on the spatial distribution of the spectrally concentrated and transitioned pixels for the generation of binary segmentation masks of the images. Avgrinak et al. \cite{avgerinakis2017visual} proposed an approach based on a Mahalanobis \cite{de2000mahalanobis} classification framework, where Mahalanobis distances with stratified covariance estimates are used for training, along with morphological operations to mitigate the false positive pixels. The authors also experiment with linear, diagonal linear, quadratic and diagonal quadratic discriminant functions. 


\ka{In MediaEval-2018 \cite{bischke2018multimediasatellite}, the participants were asked to automatically identify passable roads in flood-affected areas in satellite imagery. In total, three methods were presented for the task. In the first method Arminal et al. \cite{Armin2018multimediasatellite} divide the satellite image-patches into sub-patches each of size $50\times 50$ pixels around each of the two given points, showing the start and end of the roads. RGB histograms are then used to extract visual features for SVM-based classification of the road segments into passable and non-passable classes. In \cite{Anastasia2018multimediasatellite}, authors rely on a transfer learning method where ResNet-50, pre-trained on ImageNet, has been fine-tuned on $224\times224$ satellite image-patches. 
Said et al. \cite{nain2018multimediasatellite} analyze two different techniques based on GANs and a CNN with transfer learning based method. Due to the small size of the dataset, GANs could not perform well and experiments were abandoned. In the transfer learning based method, the authors rely on Inception V-3 ~\cite{szegedy2015rethinking}, pre-trained on the ImageNet, and the retraining method described in~\cite{donahue2014decaf}.  }

Table \ref{satellite_images_summary} provides a summary of some of the approaches for disaster analysis in satellite imagery discussed in this section. 
\begin{table*}[]
\centering
\caption{Summary of some relevant works on disaster detection in satellite images in terms of event types, dataset, and a brief description of the method.}
\label{satellite_images_summary}
\scalebox{0.69}{
\begin{tabular}{|p{.5cm}|p{2.5cm}|p{2.0cm}|p{6.0cm}|}
\hline
\multicolumn{1}{|c|}{Refs.} & \multicolumn{1}{c|}{Events} & \multicolumn{1}{c|}{Dataset} & \multicolumn{1}{c|}{Method} \\ \hline

\cite{liu2016geological} & Natural disaster events & Self-collected &  Relies on wavelet transformation in the pre-processing while a deep auto-encoder comprising of several hidden layers is used for the extraction of deep features. A softmax classifier is then used for the classification purposes.\\ \hline

\cite{bischke2016contextual} & Floods and snow-storm & A self-collected and Landsat dataset & Relies on multiple CNN models as feature descriptors with a late fusion mechanism. In addition, a crawler and a filtering scheme is proposed to retrieve relevant images from social media and LandSat dataset \\ \hline

\cite{ahmad2018social} & Floods & FDSI-2017 & Relies on a Generative Adversarial Network originally developed for vessels recognition in rental images with a threshold controlling mechanism. Provides a detailed analysis of the importance of IR component in the classification task with diverse strategies. In addition, several experiments are conducted for parameterization of the threshold values\\ \hline

\cite{nogueira2018exploiting}& Floods & FDSI-2017 & Relies on deep architectures with the concept of dilated and de-convolution  where several experiments are conducted with different number of dilated convolutional layers in the network. Moreover, the individual models based on the concept of dilated and de-convolution are combined in an ensemble framework.\\ \hline

\cite{bischke2017detection}& Floods & FDSI-2017 & Relies on an existing deep model VGGNet with three different training strategies. Moreover, a pre-processing technique is used to enhance the satellite imagery before classifying into flooded and non-flooded regions. \\ \hline

\cite{tkachenko2017wisc}& Floods & FDSI-2017 & proposes a solution based on the selection of spectral images with three indices, namely LWI (Land Water Index), NDVI (Normalised Difference Vegetation
Index) and NDWI (Normalised Difference Water Index). For the classification purposes both supervised classification and unsupervised clustering techniques are used.\\ \hline

\cite{avgerinakis2017visual}& Floods & FDSI-2017 & Relies on a Mahalanobis \cite{de2000mahalanobis} classification framework for the classification of flooded and non-flooded regions along with morphological operations to mitigate the false positive pixels. \\ \hline

\cite{kamilaris2018disaster} & Floods, fires and building collapsed & Self-collected & An existing deep model \cite{simonyan2014very} pre-trained on ImageNet \cite{deng2009imagenet} is fine-tuned on aerial photos captured through unmanned aerial vehicles (UAV) during or after different types of natural disasters \\ \hline

\cite{nain2018multimediasatellite} & Floods & FDSI-2018 \cite{bischke2018multimediasatellite} & Aims at identification and differentiating in passable and non-passable roads. Relies on Inception v-3, pretrained on ImageNet, and transfer learning method with different training strategies. Moreover, also uses data augmentation to increase the training set. Achieved best results on MediaEval 2018 challenge on FDSI task. \\ \hline

\cite{Anastasia2018multimediasatellite} & Floods & FDSI-2018 & Relies on an existing deep model namely ResNet-50, pre-trained on ImageNet, has been fine-tuned for road passability task in satellite image patches each of size $224\times224$. Other parameters are: Stochastic Gradient Descent (SGD) optimizer and a learning rate of .001 in 18 epochs. \\ \hline

\cite{Armin2018multimediasatellite} & Floods& FDSI-2018 & Mainly relies on RGB histogram features and SVM based classification of the satellite image patches, which were firstly divided into sub-patches each of size $50\times 50$ pixels around each of the two given points showing the start and end of the roads. \\ \hline

\end{tabular}}
\end{table*}

\section{\ka{Current Trends in Disasters Analysis}}

\begin{figure}[]
	\centering
	\fbox{\includegraphics[width=0.9\linewidth]{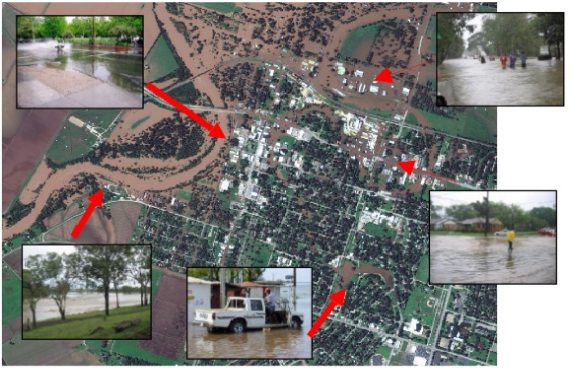}}
	\caption{Enrichment of satellite information with social media to provide a better story of natural disasters \url{https://goo.gl/1Z6E2v}. 
		\label{enrichment}}
\end{figure}

A new and not fully explored trend is to jointly use the content from social media together with satellite information. Benchmarks initiatives are emerging around this topic such as the MediaEval 2017-2018 benchmark initiative with the "Multimedia Satellite" task\footnote{\url{http://www.multimediaeval.org/mediaeval2017/multimediasatellite/index.html}}. For the task, researchers try to analyze disasters information from social media and satellite. Additionally, a task	to	build	a	system	that links	social	multimedia	to	events detected in satellite images, has been introduced as a challenge at ACM Multimedia 2016\footnote{\url{http://www.acmmm.org/2016/wp-content/uploads/2016/03/ACMMM16_GC_Sky_and_the_Social_Eye_latest.pdf}}. The basic goal of the tasks is to enrich satellite imagery with social media information, where the satellite imagery provides a big picture of the disaster, while social media is intended to provide more localized and detailed information about the disaster in the form of text, images, and videos, as shown in Figure \ref{enrichment}. 

The first effort in this direction is made in \cite{bischke2016contextual}, where a system for the contextual enrichment of satellite imagery has been proposed by collecting and analyzing multimedia content (text, images and videos) from Twitter. The system initially collects satellite imagery from LandSat\footnote{https://landsat.usgs.gov/landsat-data-access} using manual queries as well as  geo-location information extracted from the names of the places mentioned in Twitter text or the GPS annotation of tweets directly, if available. A crawler has also been developed to collect natural disaster-related multimedia content from Twitter using specific keywords. The collected multimedia content is then processed and analyzed, individually, to filter out irrelevant information. The geo-location information associated with multimedia content is then used to map the multimedia content on the satellite imagery. Moreover, a user interface has been provided with different visualization techniques to facilitate the potential users for exploring and extracting information for disaster events.

In \cite{ahmad2018social}, a system called JORD has been introduced to collect, filter and link information form social media and satellites. As a first step, JORD crawls disaster occurrence information from  EM-DAT\footnote{https://www.emdat.be} and starts collecting information from several social media platforms with queries in the different languages spoken in the affected areas. Subsequently, the retrieved information is processed and analyzed. Satellite imagery is crawled using the geo-location information obtained from the retrieved multimedia content (TWitter text, images and videos) and EM-DAT. Moreover, Twitter text is processed and analyzed to extract the names of places and cities, which are then used for collecting satellite imagery. As a final step, a user interface is developed to show a joint view of an underlying disaster containing the multimedia content from social media and satellite imagery. In addition, as a test-used case, a GANs based approach has been proposed for flood detection in satellite imagery. 


\section{Datasets}
Statistics and other information of the damage, are crucial to understand the potential impact of a disaster on the affected areas as well as the cost associated with the recovery process. To this aim, there are a number of organizations, who systematically collect and provide data and statistics of the disasters occurred all over the world. These databases vary in terms of type of information and geographical area, resolution and the methodologies adopted for the data collection. The databases can be divided into four different categories, namely (i) databases on disaster loss and damage, (ii) hazard catalogues, (iii) socio-economic indicators and (iv) exposure datasets\footnote{\url{https://www.preventionweb.net/risk/datasets\#panel1-4}}. The databases on the disaster loss and damage contain statistics of the disasters occurred in the past, and allow us to explore and understand the trends of risks associated with different disaster types. For instances, EM DAT \cite{guha2015dat} and DesInventar \cite{unisdr2011robot} provide detailed statistics of the different natural and technological disasters occurred world wide. The Hazard catalogues, such as Advance National Seismic System (ANSS) \cite{herrmann2008april} and GEM \cite{pinho2012gem}, maintain a record of basic information about disasters, such as the occurring date, geographical location, scope, magnitude and intensity. The socio-economic indicators, such as World development indicators \cite{world1999world}, and Exposure datasets, such as GED4GEM \cite{gamba2012ged4gem}, help to assess and monitor socio-economic vulnerability and resilience to disasters.

\ka{The datasets mentioned above are, however, rather generic and contain the historical data about the occurred events. In our discussion we prefer instead to focus our attention on very specific information sources that can be adopted for benchmarking, with the aim of developing algorithms for disaster detection. }

\subsection{Twitter Datasets}

CrisisLex\footnote{\url{http://crisislex.org/data-collections.html\#CrisisLexT26}}, which is an online repository, provides a platform for sharing crisis-related data collections and other tools to analyze social media data for disasters. Several collections of disaster-related tweets have been provided on the platform. For instance, CrisisLexT26 \cite{olteanu2015expect} is a collection of 250,000 tweets, gathered during 26 different crisis events happened in 2012 and 2013. The dataset provides on the average from 2,000 to 4,000 tweets per crisis event. Tweets are annotated through a crowd-sourcing activity, where around 28,000 tweets are annotated on the basis of informativeness (i.e., informative vs. non-informative), information type (i.e., caution and advice, infrastructure damage), and sources of information (i.e., governments, NGOs). The dataset is provided in CSV file containing Twitter ID, label and text of the tweets along with meta-data. CrisisLexT6 \cite{olteanu2014crisislex} is another collection of crisis events-related tweets provided by the same authors. The dataset covers 6 different types of crisis events, namely 2012 Sandy Hurricane, 2013 Boston Bombings, 2013 Oklahoma Tornado, 2013 West Texas Explosion, 2013 Alberta Floods and 2013 Queensland Floods. In total, it includes 60,000 tweets annotated via crowd-sourcing on the basis of relatedness (i.e., on-topic vs off-topic).

ChileEarthquakeT1 \cite{cobo2015identifying}, provides a collection of 2,187 tweets in Spanish language, which were posted during the Chilean earthquake in 2010. Tweets are labelled on the basis of relatedness (true if related otherwise false) using six different annotators with at least three annotators per tweet. SoSItalyT4 \cite{cresci2015linguistically} provides a collection of 5,600 tweets posted during 2009 to 2014, giving information about 4 natural disasters (2 earthquakes and 2 floods) in Italy. The dataset is annotated with three annotators with damage and no damage labels. Wang et al. \cite{wang2015hurricane} provide another dataset, namely SandyHurricaneGeoT1, composed of geo-tagged tweets from 2012 Sandy hurricane. A total of 6,556,328 geo-tagged tweets representing the time and regions impacted by the hurricane. The dataset is provided as a CSV file with tweet ID, time-stamps and geo-location information. However, no labels are provided.

\ka{More recently, Olteanu et al. \cite{olteanu2015comparing} published ClimateCovE350, a collection of tweets from around 350 climate-related events. These events cover a wide range of emergency situations ranging from natural disasters and hazards, such as typhoon, tornadoes and droughts, to human induced hazards, such as deforestation and oil spill. The dataset is labelled with a diverse set of tags including labels on the basis of relevance to climate-chance, triggers, actions, and six news values (i.e. extraordinary, unpredictable, high magnitude, negative, conflictive, related to elite persons). McMinn et al. \cite{mcminn2013building} collected a large-scale event detection corpus. It covers a large set of events including business, economic, education, arts as well as disasters and accidents. In total, the dataset covers 510 Twitter events including twenty nine disasters and accidental events. In \cite{imran2013extracting}, a collection of tweets from Joplin tornado has been provided. The tweets are categorized and annotated based on the humanitarian categories including caution and advice, donations, information sources and casualties. Another dataset composed of tweets and associated images related to seven different natural disaster events, namely Hurricane Irma, Hurricane Harvey, Hurricane Maria, Mexico earthquake, California fires, Iraq-Iran earthquake and Srilanka floods. The dataset is intended to support three different tasks including informative and non-informative disaster content, humanitarian categories and damage severity assesment. }

Table \ref{dataset_tweets} summarizes the properties of the most relevant datasets discussed in this sub-section.

\begin{table*}[]
\centering
\caption{\ka{Summary of the benchmark datasets for natural disasters detection in Twitter.}}
\label{dataset_tweets}
\scalebox{0.9}{
\begin{tabular}{|p{.2cm}|p{1.3cm}|p{2.2cm}|p{2.5cm}|p{.2cm}|p{4cm}|}
\hline
\multicolumn{1}{|c|}{\textbf{Refs.}} & \multicolumn{1}{|c|}{\textbf{Sampling}} &\multicolumn{1}{|c|}{\textbf{Name}}& \multicolumn{1}{c|}{\textbf{Features}} & \multicolumn{1}{c|}{\textbf{A}}& \multicolumn{1}{c|}{\textbf{Comments}} \\ \hline \hline

\cite{olteanu2015expect} & keyword-based queries & CrisisLexT26 & Two classes and 250K tweets & Y & Publicly available collection of tweets during twenty six large crisis events in 2012 and 2013. \\ \hline 

\cite{olteanu2014crisislex} & keyword and geo-based queries & CrisisLexT6 & Two classes and 60,000 tweets & Y & Publicly available collection of tweets collected during six large crisis events in 2012 and 2013. \\ \hline 

\cite{cobo2015identifying} & keyword-based queries & ChileEarthquakeT1 & Two classes and 2,100 K tweets & Y & Publicly available collection of tweets collected during 2010 Chilean earthquake. Avaliable in a CSV file with tweet ID, text and corresponding labels.\\ \hline 

\cite{cresci2015linguistically} & keyword-based queries & SoSItalyT4 & Three classes and 5,600 K tweets & Y & Provides tweet related to four events (two floods and two earthquakes) accursed in Italy between 2009 to 2014. Available in a CSV file with tweet ID, text and corresponding labels.\\ \hline

\cite{wang2015hurricane} & geo-based queries & SandyHurricaneGeo & two classes (related vs. not related) & Y & Provides a total of 6,556,328 geo-tagged tweets representing the time and regions impacted by the hurricane. Available in CSV file with Tweet ID, time-stamps and geo-location information. \\ \hline 

 \cite{olteanu2015comparing} & keyword-based queries &  ClimateCovE350 & Nine different classes& Y & Provides a collection of tweets from around 350 climate related events. Covers a wide range of emergency situations ranging from natural disasters and hazards, such as typhoon, tornadoes and droughts, to human induced hazards, such as deforestation and oil spill.  \\ \hline 

 \cite{crisismmd2018icwsm} & keyword-based queries &  CrisisMMD & Tweets and images from three different types of disasters  labeled for three different tasks& Y & Provides a collection of tweets and associated images from seven disaster events including three hurricanes, two earthquake and one for fire and flood. The tweets and images are annotated for three different types of tasks each one with different number and types of labels. \\ \hline 

\end{tabular}}
\end{table*}

\subsection{Images Datasets}
Considering the growing interest of the research community in content analysis of disaster-related images from social media, different benchmarking activities have been organized for the evaluation and comparison of disaster detection frameworks in single images. In this regard, the most widely used dataset is Disaster Image Retrieval from Social Media (DIRSM) \cite{bischke2017multimedia} provided in MediaEval 2017 benchmarking initiative task on multimedia and satellite. The dataset is composed of two classes, namely flooded and non-flooded images, and is provided in development and test sets containing 5,280 and 1,320 images obtained from YFCC100M \cite{thomee2016yfcc100m}, respectively. The dataset also provides meta-data from YFCC100M along with visual global features extracted through Lire library \cite{lux2016lire} for each image.

In MediaEval 2018 \cite{bischke2018multimediasatellite}, a benchmark dataset for flood classification of images providing an evidence for road passability in social media has been provided. The dataset contains a total of 11,070 images along with the associated meta-data from three different hurricane events, namely Harvey, Irma and Maria, occurred in 2017. The dataset has been provided in a development set containing 7,387 and a test set composed of 3,683 images. Since the dataset is provided for two sub-tasks i.e., (i) identification of images providing an evidence of passability and (ii) differentiating passable and non-passable roads in images, ground truth has been provided as two separate files. Moreover, a collection of the corresponding visual features/descriptors for each image extracted through several algorithms has been also provided.

Table \ref{dataset_images} provides a summary of the properties of the datasets discussed in this sub-section.
\begin{table*}[]
\centering
\caption{Summary of the benchmark datasets for natural disasters detection in images from social media. \ka{All the datasets are annotated and "B" indicates whether the dataset has been part of a benchmark competition or not}}
\label{dataset_images}
\scalebox{0.8}{
\begin{tabular}{|p{.3cm}|p{1.6cm}|p{2.2cm}|p{2.5cm}|p{.3cm}|p{5cm}|}
\hline
\multicolumn{1}{|c|}{\textbf{Refs.}} & \multicolumn{1}{|c|}{\textbf{Modality}} &\multicolumn{1}{|c|}{\textbf{Name}}& \multicolumn{1}{c|}{\textbf{Features}} & \multicolumn{1}{c|}{\textbf{\ka{B}}}& \multicolumn{1}{c|}{\textbf{Comments}} \\ \hline \hline

\cite{bischke2017multimedia}& Images and meta-data  & DIRSM & Two classes and 6,600 images & Y & The first publicly available dataset for disaster analysis in images; covers floods events only; Meta-data composed of user's tags, temporal, geo-location and owner information are also provided; moreover, global features extracted through several feature descriptors \cite{lux2016lire} are provided to facilitate the participants\\ \hline

\cite{bischke2018multimediasatellite}& Images and meta-data & FCSM & Two classes  and 11,070 images & Y & covers flood related images from three different hurricanes occurred in 2017; Also accompanied by meta-data including users' tags, textual description as well as rainfall and climate prediction; visual features extracted through algorithms are also provided; covers two different sub-tasks; Manually annotated and filtered. \\ \hline

\end{tabular}}
\end{table*}

\subsection{Satellite Imagery Datasets}

Although disaster detection in satellite imagery got a relevant amount of attention of the research community, and a number of interesting solutions have been proposed, a proper dataset for the evaluation and comparison is still lagging behind. There are a number of reasons for the unavailability of such datasets. Some of the important factors include the commercial aspects/use of the satellite imagery, and low temporal frequency of the satellite imagery.

NASA has recently released the data of Landsat\footnote{https://landsat.usgs.gov/landsat-8}, the longest running satellite program that has opened a number of opportunities for the researchers, enabling them to develop systems that integrate remotely sensed data in different applications. The literature shows the use of satellite imagery obtained from Landsat archives in several works for disaster analysis \cite{fisher2016comparing,kansas2016using}. However, to the best of our knowledge, the majority of the works use their own set of images for the evaluation purposes, which are not available for the comparison purposes.

In MediaEval 2017, a benchmarking dataset namely Flood Detection in Satellite Imagery 
(FDSI) has been provided. The dataset contains satellite image-patches obtained from the Planet's 4-band satellites with ground-sample distance (GSD) of 3.7 meters \cite{team2016planet}. The dataset mainly covers images from floods obtained during or after 8 different flood events from 01.06.2016 to 01.05.2017. The image-patches are provided in 4 channels (i.e., RGB and IR component). For the challenge, the dataset is provided into development and test tests. The development set contains a total of 462 image-patches from six different locations. However, the test set has been further divided into two subsets. The first test set contains unseen image patches from the same locations covered in the development set, while the test set 2 contains unseen image patches from different location not covered in the development dataset. 


In \cite{bischke2018multimediasatellite}, a dataset is provided for detection of passable and non-passable roads in satellite imagery in the MediaEval 2018 benchmark initiative. In total, the dataset contains 1,663 image-patches provided by DigitalGlob\footnote{https://www.digitalglobe.com}. The images are divided into development and test sets containing 1,438 and 225 image-patches, respectively. The annotations are provided in binary form i.e., passable and non-passable road segments. 

In Table \ref{dataset_satellite}, we provide a summary of the datasets available for the disaster events detection and analysis in satellite imagery.
\begin{table*}[]
\centering
\caption{Summary of the datasets for the analysis of natural disaster related multimedia content (text from Twitter, images from Flickr and other platforms of social media, and satellite images. \ka{All the datasets are annotated and "B" indicates whether the dataset has been part of a benchmark competition or not.}}
\label{dataset_satellite}
\scalebox{0.8}{
\begin{tabular}{|p{.3cm}|p{1.0cm}|p{2.2cm}|p{2.5cm}|p{.3cm}|p{5.6cm}|}
\hline
\multicolumn{1}{|c|}{\textbf{Refs.}} & \multicolumn{1}{|c|}{\textbf{Data Type}} &\multicolumn{1}{|c|}{\textbf{Name}}& \multicolumn{1}{c|}{\textbf{Features}} & \multicolumn{1}{c|}{\textbf{\ka{B}}}& \multicolumn{1}{c|}{\textbf{Comments}} \\ \hline \hline

\cite{bischke2017multimedia}& Satellite images & FDDI-2017 & Two classes and around 700 image patches & Y & The first publicly available dataset for disaster analysis in satellite  images; covers floods events only; image patches are provided from 8 different locations; image patches are provided in 4-channels, namely R, G, B and IR; Also provides temporal information, which can help in time based analysis for flood detection in satellite imagery \\ \hline



\cite{bischke2018multimediasatellite} & Satellite images & FDSI-2018 & Two classes and 1,663 image patches & Y & Mainly covers images from three different hurricanes, namely Irma, Harvey, and Maria, occurred in 2017. Annotations are provided in binary form for the whole image patch. Each image patch has a spatial resolution of $512 \times 512$, and covers ground-sample distance of 0.5 meters.\\ \hline 

\end{tabular}}
\end{table*}

\section{Summary and future directions}
In this survey paper, we conducted a comprehensive analysis of different approaches proposed for disaster detection, retrieval, summarization and analysis of the retrieved information from different platforms of social media as well as satellite imagery. We have discussed several approaches for disasters analysis in three different sub-domains, namely disaster detection and analysis in Twitter text streams, disaster detection in visual content (images and videos) available in social media, and detection of disaster's affected areas in satellite imagery. We also discussed the current trends, benchmarking initiatives and challenges associated with disaster analysis in theses sub-domains. Being one of the most important part of the literature, we also discussed about the benchmarking datasets available for evaluation.

The literature shows that Twitter has been explored extensively for disaster detection over the years. The majority of the approaches rely on neural networks with diverse strategies to analyze Twitter text streams. User-captured images from social media have also been widely analyzed. We observed a trend towards the use of deep architectures. The majority of the approaches rely on existing pre-trained models, which are either fine-tuned on disaster images or used as feature descriptors. In this regard, the models pre-trained on ImageNet and places datasets have been heavily exploited individually and in combination. We observed that in contrast to other categories of events (e.g., social and sports) in disaster events detection and analysis the models pre-trained on places datasets, which corresponds to scene-level information, tend to perform better compared to the ones pre-trained on ImageNet. In terms of performance, the fusion of different CNN models outperformed individual models. Moreover, the additional information available in the form of meta-data have also been widely utilized to support visual information. Satellite imagery has also been proven as one of the most useful sources of information for disaster monitoring and analysis to estimate the damage and cost associated with them. Similar to disaster detection in images from social media, the majority of the approaches proposed for disaster detection in satellite imagery relied on CNNs with a significant improvement over traditional handcrafted visual features. 


\ka{We also provide the readers with an overview of the state-of-the-art performances in quantitative terms on the most widely used datasets in the above-discussed research areas. Though the numbers are expected to grow, we aim to set a marker line according to the existing literature to be used as a reference for future works. In terms of disaster events detection and analysis in Twitter text streams, the majority of the proposed approaches used self-collected datasets for the evaluation, which makes the quantitative comparison of the methods very difficult. However, in the case of image-based analysis we show that the datasets provided in benchmark initiatives are a good way to quantitatively compare different methods. As far the natural disaster analysis in images from social media is concerned, the average accuracy reached by the relevant methods \cite{ahmad2018comparative,ahmad2017cnn,bischke2017detection,nain2018multimediasatellite,Anastasia2018multimediasatellite} on benchmark DIRSM \cite{bischke2017multimedia} and FCSM \cite{bischke2018multimediasatellite} datasets is around 95\% and 65\%, respectively, showing significantly higher complexity for the FCSM dataset. On the other side, in the case of disaster detection in satellite imagery, the performances are on the lower side where the methods proposed in \cite{ahmad2017cnn,bischke2017detection,nogueiradata,nain2018multimediasatellite,Armin2018multimediasatellite} could achieve a mean F-score of 65\% and 62\% on the widespread datasets FDSI-2017 and FDSI-2018, respectively.}


\ka{As described in detail in Section 2, the literature on disaster analysis lacks large-scale benchmark datasets, in terms of both total number of images and the types of natural disasters covered. For instance, DIRSM \cite{bischke2017detection} and FCSM \cite{bischke2018multimediasatellite} cover only flood events and include a relatively small number of samples. In the case of satellite imagery, the problem is even worst where very few annotated image-patches are publicly available and, again, only for floods detection. Collecting large-scale benchmark datasets covering different types of natural disaster events with multi-modal information is one of the aspects that need to be further explored. Moreover, the authenticity of the retrieved information from social media is also an important aspect to be researched.}

\ka{To the best of our knowledge, most of the existing literature relies on images only for visual content based analysis of natural disasters. Another interesting direction of future work could be the content analysis of natural disaster-related videos. Although more challenging, videos could provide more detailed information and are expected to improve users' experience. }

\ka{We also believe that the additional sources of information, such as OSM and other GIS datasets, along with satellite imagery can further improve the performances of AI techniques in disaster analysis. 
}

\ka{Another encouraging aspect is the availability of statistics and other information about the natural disasters, which can help to understand the potential impact of a disaster on the affected areas as well as the cost associated with the recovery process. Such databases containing statistics of the disasters occurred in the past can be explored to understand and predict the trends and risks associated with different disaster types in different parts of the world. An integration and automatic analysis of such sources of information could be an interesting direction for future research in the domain.}


{\small
\bibliographystyle{ieee}
\bibliography{egbib}
}

\end{document}